\documentclass{article}

\usepackage{subfigure}
\usepackage{natbib}
\usepackage{amsmath}
\usepackage{amssymb}
\usepackage{graphicx}
\usepackage{latexsym}
\usepackage{aas_macros}
\usepackage{fancyhdr}
\usepackage{lastpage}

\usepackage{url}

\usepackage[squaren]{SIunits}

\newcommand{\scinot}[2]{\ensuremath{#1 \times 10^{#2}}}

\newcommand{\Mstar}[1]{\ensuremath{M_{*}^{#1}}}
\newcommand{\Mplanet}[1]{\ensuremath{M_{\text{p}}^{#1}}}

\addunit{\cm}{\centi\metre}

\addunit{\dyne}{dyn}
\addunit{\erg}{erg}

\addunit{\gauss}{gauss}

\addunit{\persqcm}{\per \cm \squared}
\addunit{\persqcmnp}{\cm \rpsquared}

\addunit{\percubiccm}{\per \cm \cubed}
\addunit{\percubiccmnp}{\cm \rpcubed}

\addunit{\grampersqcm}{\gram \persqcm}
\addunit{\grampersqcmnp}{\gram \usk \persqcmnp}
\addunit{\grampercubiccm}{\gram \percubiccm}
\addunit{\grampercubiccmnp}{\gram \usk \percubiccmnp}

\addunit{\ergpercubiccm}{\erg \percubiccm}
\addunit{\ergpercubiccmnp}{\erg \usk \percubiccmnp}

\addunit{\molpercubiccm}{\mole \percubiccm}
\addunit{\molpercubiccmnp}{\mole \usk \percubiccmnp}

\addunit{\cmpersec}{\cm \per \second}
\addunit{\cmpersecnp}{\cm \usk \reciprocal \second}
\addunit{\kmpersecnp}{\kilo\metre\usk\reciprocal\second}

\addunit{\cmpersecsq}{\cm \per \second \squared}
\addunit{\cmpersecsqnp}{\cm \usk \second \rpsquared}

\addunit{\gramcmpersec}{\gram \usk \cmpersec}
\addunit{\gramcmpersecnp}{\gram \usk \cmpersecnp}

\addunit{\gramsqcmpersec}{\gram \usk \cm \squared \per \second}
\addunit{\gramsqcmpersecnp}{\gram \usk \cm \squared \second \rpsquared}

\addunit{\yyear}{yr} 
\addunit{\MBU}{MBU} 

\addunit{\jansky}{Jy}

\addunit{\magnitude}{mag}

\addunit{\cmsqpergramnp}{\centi\metre\squared\usk\reciprocal\gram}

\newcommand{\mySun}{\odot}

\addunit{\Msol}{\ensuremath{\mathrm{M}_{\mySun}}}
\addunit{\Rsol}{\ensuremath{\mathrm{R}_{\mySun}}}
\addunit{\Lsol}{\ensuremath{\mathrm{L}_{\mySun}}}
\addunit{\Zsol}{\ensuremath{\mathrm{Z}_{\mySun}}}

\providecommand{\earth}{\oplus}

\addunit{\Mearth}{\ensuremath{\mathrm{M}_{\earth}}}
\addunit{\Rearth}{\ensuremath{\mathrm{R}_{\earth}}}

\addunit{\Mjup}{\ensuremath{\mathrm{M}_{J}}}
\addunit{\Rjup}{\ensuremath{\mathrm{R}_{J}}}

\addunit{\AU}{au}
\addunit{\lightyear}{ly}
\addunit{\parsec}{pc}

\newcommand{\qdisc}[1]{\ensuremath{q_{\textrm{disc}}^{#1}}}

\begin{document}

\title{Type II Migration: Varying Planet Mass and Disc Viscosity}
\author{R.~G.~Edgar}
\date{\today}

\maketitle


\abstract{This paper continues an earlier study of giant planet migration, examining the effect of planet mass and disc viscosity on the migration rate.
We find that the migration rate of a gap-opening planet varies systematically with the planet's mass, as predicted in our earlier work.
However, the variation with disc viscosity appears to be much weaker than expected.}


\section{Introduction}

This work is a continuation of \citet{2007ApJ...663.1325E}, which showed that Type II migration (the migration of a gap-opening planet) was not as simply as commonly thought.
Most previous workers assumed that, once a planet opened a gap in a disc, it would migrate along with the general viscous evolution of the disc itself.
One particular prediction of this approach is that the migration rate of the planet should only vary with the disc viscosity.
The disc mass should not affect the migration rate, unless the disc mass were very low \citep[][ examined this limit]{1995MNRAS.277..758S,1999MNRAS.307...79I}.

\citet{2007ApJ...663.1325E} demonstrated that the migration rate of a gap-opening planet was directly proportional to the disc surface density, in contradiction of the usual assumption.
The planet is not a test particle within the disc, but must exchange angular momentum with it.
As a result, the migration rate will continue to depend on the various masses involved.

\cite{2004ApJ...604..388I} suggested an alternative prescription for determining the Type II migration rate.
They balanced the angular momentum change of the planet with the maximum viscous couple in the disc.
Following this idea, \citet{2007ApJ...663.1325E} predicted that the Type II migration rate should be
\begin{equation}
\dot{a} = - 3 \frac{\nu \Sigma a}{\Mplanet{}}
\label{eq:TypeIIAngMomBalance}
\end{equation}
where the planet's semi-major axis is $a$, the local disc surface density is $\Sigma$, this gas viscosity is $\nu$ and the planet's mass is \Mplanet{}.
In this paper, we shall study the variation in migration rate of a gap-opening planet with the planet's mass and the disc viscosity.

\section{Numerical Details}
\label{sec:numeric}

Our basic technique is identical to that of \citet{2007ApJ...663.1325E}, using the \textsc{Fargo} code of \citet{2000A&AS..141..165M,2000ASPC..219...75M}.
\textsc{Fargo} is a simple \textsc{2d} polar mesh code dedicated to disc planet interactions. 
It is based upon a standard, \textsc{Zeus}-like \citep{1992ApJS...80..753S} hydrodynamic solver, but owes its name to the \textsc{Fargo} algorithm upon which the azimuthal advection is based.
This algorithm avoids the restrictive timestep typically imposed by the rapidly rotating inner regions of the disc, by permitting each annulus of cells to rotate at its local Keplerian velocity and stitching the results together again at the end of the timestep.
The mesh centre lies at the central star, so indirect terms coming from the planets and the disc are included in the potential calculation.
We make use of a non-reflecting inner boundary, to prevent reflected waves from interfering with the calculations.
The pitch angle of the wake is evaluated in the WKB approximation.
The inner ring of active cells is then copied to the ghost cells, with an azimuthal shift appropriate to the pitch angle.
Material which flows off the inner boundary is not added to the star (nor does the planet itself accrete).
At the outer boundary, mass was added, to compensate for the viscous evolution of the system.

We use units normalised such that $G=\Mstar{}+\Mplanet{}=1$, while the planet's initial orbital radius is set at $a=1$.
References to times in terms of `orbits' should be understood to mean ``orbital times at the planet's initial radius.''
The grid extends between $r=0.4$ and $r=2.5$.
Scaled to Jupiter's orbit, this grid roughly covers the area between the asteroid belt and Saturn's orbit.
We assume a constant aspect ratio disc, with $h/r = 0.05$, and initially constant surface density, $\Sigma_0$.
We set $\Sigma_0$ through
\begin{equation}
\qdisc{} = \frac{\Sigma_0 \pi a^2}{\Mstar{}}
\label{eq:qDiskDefine}
\end{equation}
which provides a quick \emph{estimate} of the disc's mass within the planet's orbit.
We set $\qdisc{} = \scinot{4}{-3}$, which ensures that our disc is always significantly more massive than the planets we consider.

The gravitational effect of the planet on the disc is smoothed at 0.6 of the disc thickness at the planet's orbital radius:
\begin{equation}
\phi = - \frac{G M}{\sqrt{r^2 + \epsilon^2}}
\end{equation}
where $\epsilon = 0.6 h$.
There are two motivations for this, the first being the desire to avoid having a singularity wandering around the grid.
The second is physical.
The \textsc{2d} approximation becomes poor close to the planet, where the vertical distribution of material becomes important.
The actual distance of material from the planet ceases to be the well approximated by the in-plane distance, which would lead to the gravitational effect being over-estimated.
Accordingly, we soften the potential over distances comparable to the disc scale height.
However, this softening length is still substantially smaller than the expected gap size and the planet's Hill sphere.
When calculating the torque the disc exerts on the planet, material from within the Hill sphere is subject to an exponential cut off, for similar reasons.
At the start of each run, the planet is introduced gradually (over about one orbit), and is not initially permitted to migrate.
This is done to minimise the effect of transients caused by the sudden appearance of a planet in a smooth disc.
We released the planet to migrate after 1000 orbits.
The computational grid is covered by 128 radial and 384 azimuthal cells (all uniformly spaced).

So far as possible, this setup mirrors that used in the comparison project presented by \citet{2006MNRAS.370..529D}.
In that comparison, the \textsc{Fargo} code was seen to give similar results to other codes used to study the disc--planet interaction problem.

\section{Results}
\label{sec:results}

We will now show the results of our runs.
We considered planet masses of 1.0, 1.5, 2.0, 2.5 and \unit{3}{\Mjup} (recall that $\Mjup / \Msol \approx 10^{-3}$) and constant gas viscosities of \scinot{2}{-6}, \scinot{8}{-6}, \scinot{9}{-6} and $10^{-5}$ in our units.
These values ensure that the planets are always gap-opening.

\begin{figure}
\begin{tabular}{cc}
\includegraphics[scale=0.6]{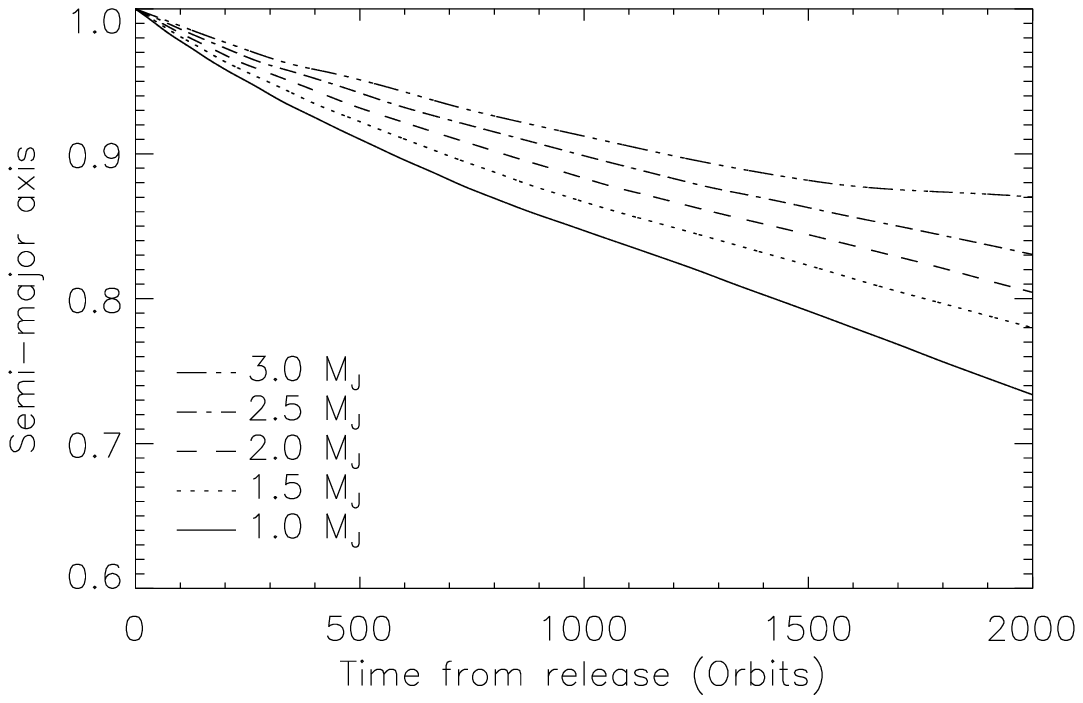} &
\includegraphics[scale=0.6]{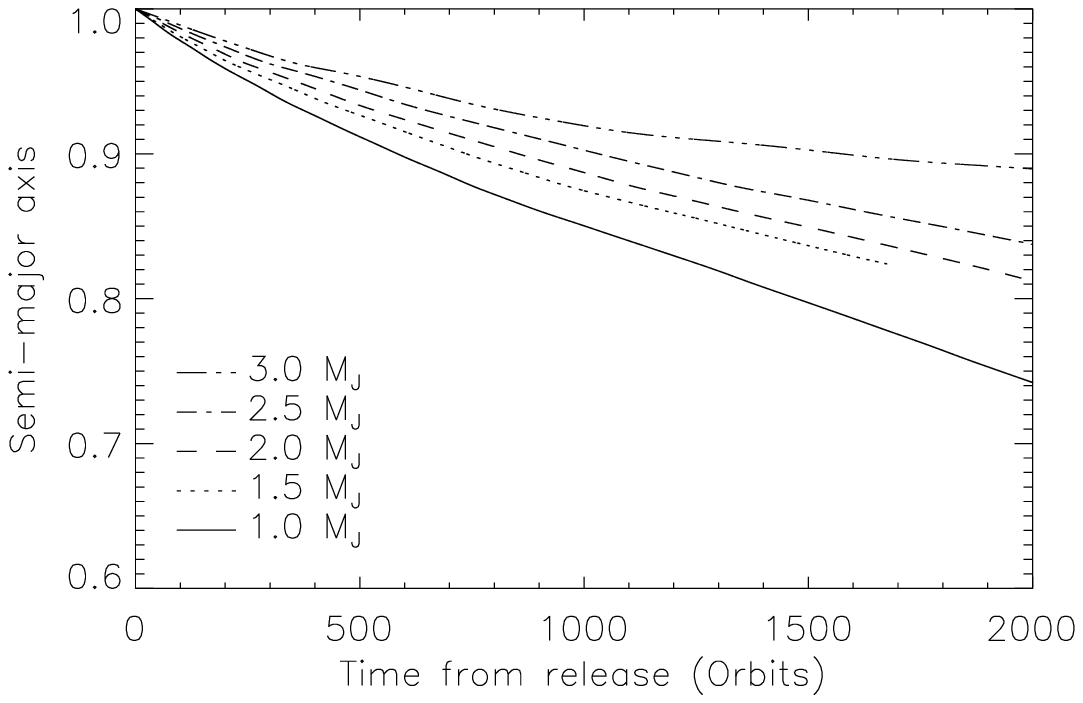} \\
\includegraphics[scale=0.6]{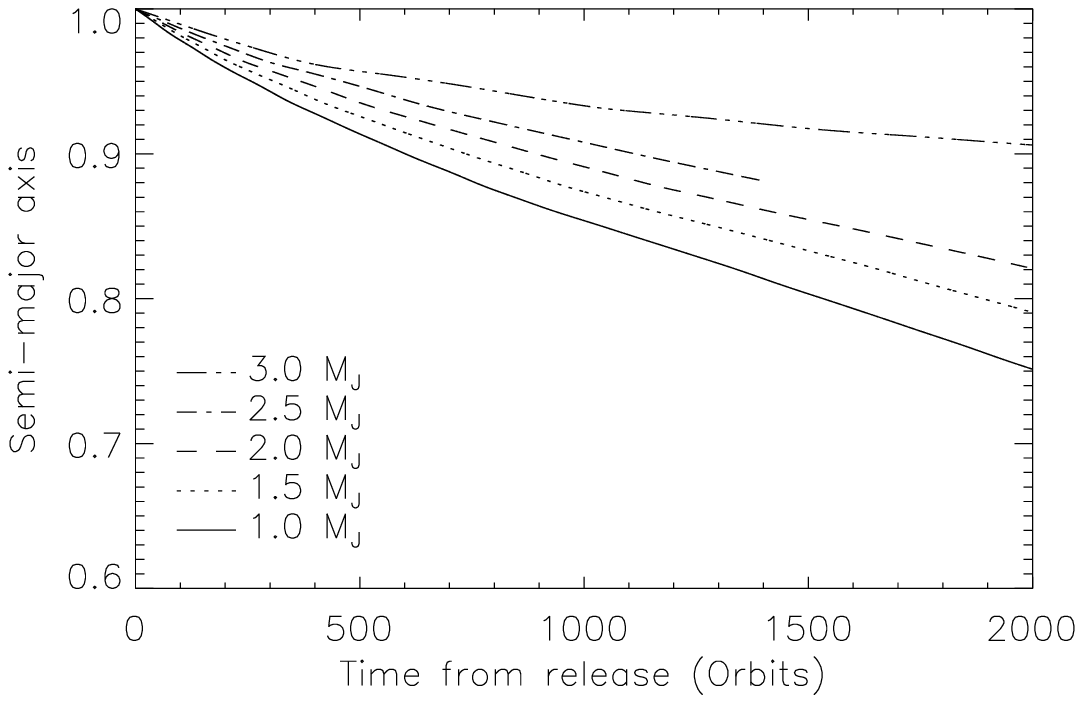} &
\includegraphics[scale=0.6]{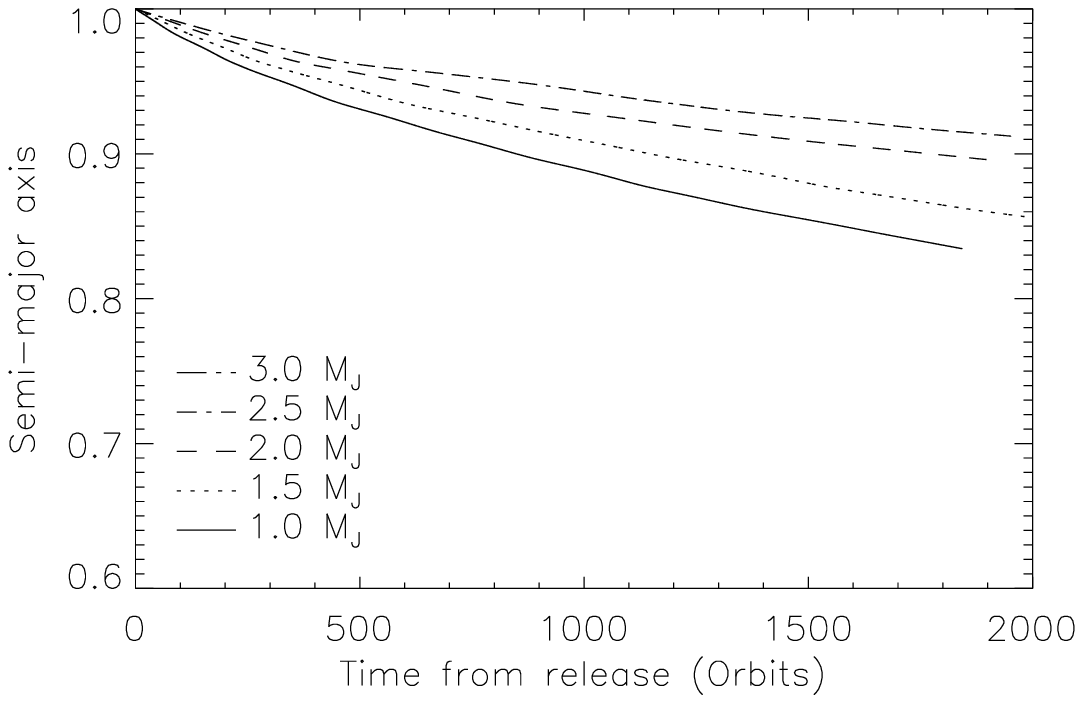}
\end{tabular}
\caption{Planetary migration in discs with constant viscosity and initially constant surface density. The lines are marked with the planet mass in each case. Each plot corresponds to a different value for the disc's viscosity: $10^{-5}$ (TL), \scinot{9}{-6} (TR), \scinot{8}{-6} (BL) and \scinot{2}{-6} (BR).}
\label{fig:MigrationRatesVaryPlanet}
\end{figure}

In Figure~\ref{fig:MigrationRatesVaryPlanet}, we show how the migration rates vary with the planet's mass in discs of differing viscosity.
Although the planets start to migrate at $t=0$ on these plots, recall that they had already completed 1000 fixed orbits, giving time for the disc to adjust to the presence of the planet.
We cut the $y$-axis at $0.6$, since at that point the $m=2$ ILR of the planet encounters the edge of the grid.
The migration rate of the planet therefore becomes unreliable.

The general pattern is clear: increasing the planet's mass decreases the migration rate.
This happens for all values of the disc viscosity, in accordance with the prediction of Equation~\ref{eq:TypeIIAngMomBalance}.
However, the value of the scaling constant is not necessarily the same.
Integrating Equation~\ref{eq:TypeIIAngMomBalance}, we find that it predicts a migration rate approximately 25\% slower than we see here.
One should compare with \citet{2007ApJ...663.1325E}, where the initial surface density profile was varied.
The variation altered the detailed migration rates, although by less than the variation in the disc mass (which was varied in that paper).

\begin{figure}
\begin{center}
\includegraphics[scale=0.6]{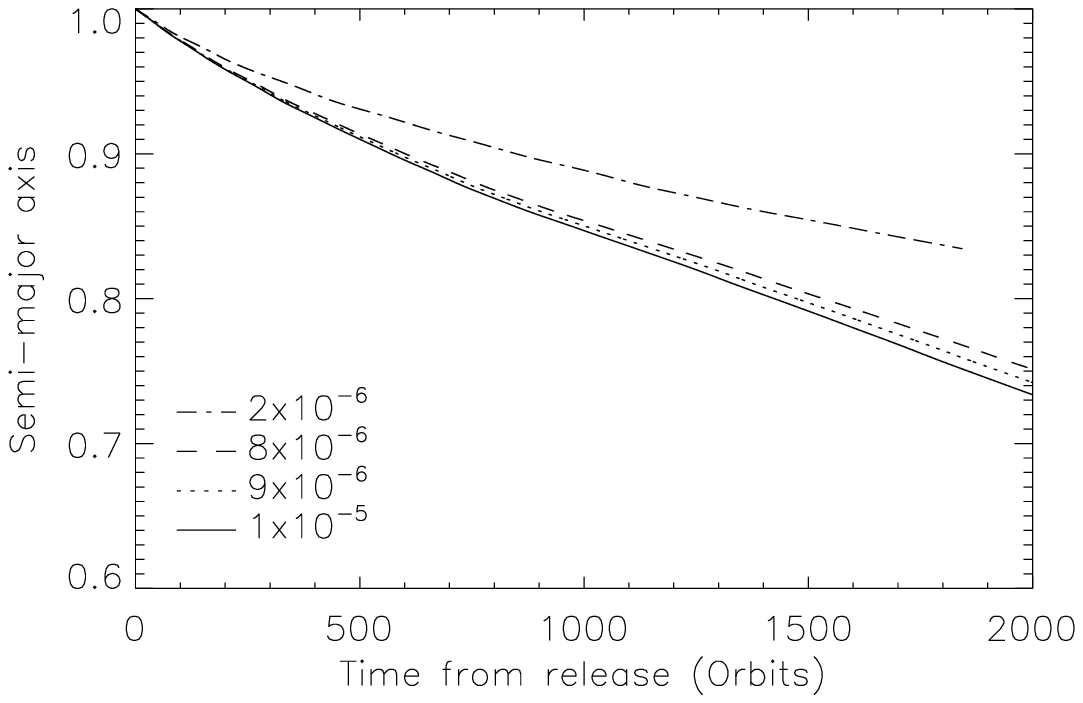} \\
\includegraphics[scale=0.6]{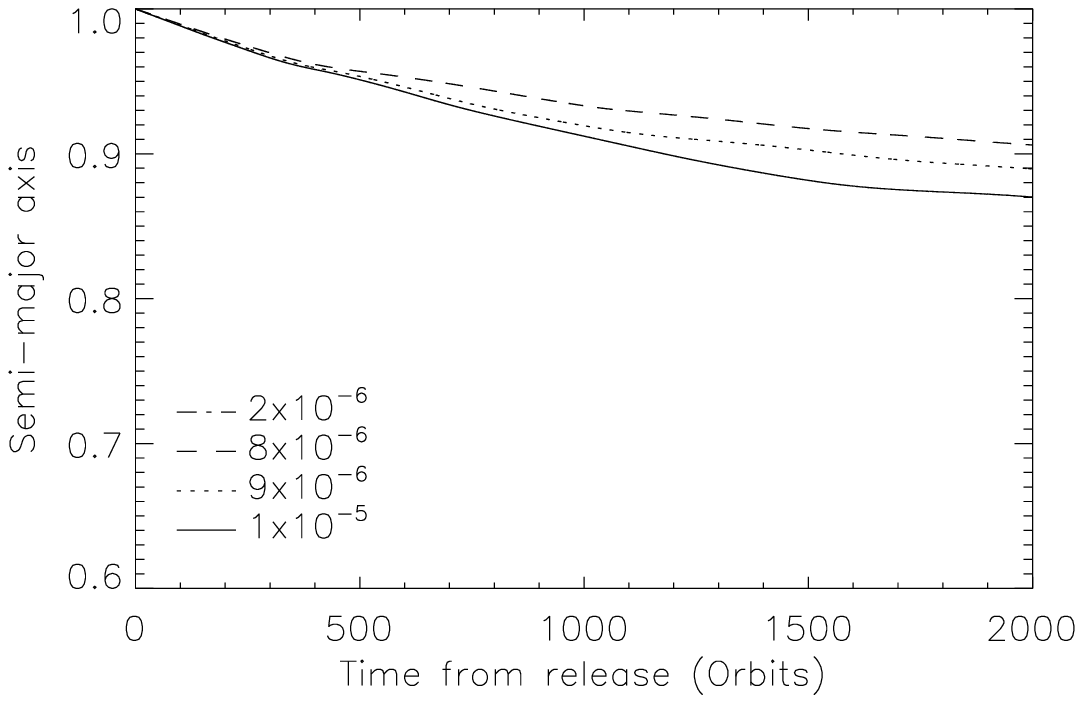}
\end{center}
\caption{Planetary migration in discs with constant viscosity and initially constant surface density. The lines are marked with the disc's viscosity. Top plot is for a \unit{1}{\Mjup} planet, the bottom for a \unit{3}{\Mjup} planet. These are the same data plotted in Figure~\ref{fig:MigrationRatesVaryPlanet}}
\label{fig:MigrationRatesVaryVisc}
\end{figure}

We will now study how the migration rate varies with the disc viscosity.
We show some sample results in Figure~\ref{fig:MigrationRatesVaryVisc}.
The data are the same as those used in Figure~\ref{fig:MigrationRatesVaryPlanet}, but plotted in different groupings.
We only show the results for the \unit{1}{\Mjup} and \unit{3}{\Mjup} cases; the other planet masses gave similar results.

We see immediately that the Equation~\ref{eq:TypeIIAngMomBalance} is not predicting the variation with disc viscosity shown by our numerical experiments.
The variation is far weaker in our runs than predicted above.
The difference cannot be due to the simplifications made in deriving Equation~\ref{eq:TypeIIAngMomBalance}, since that should not affect the scaling with $\nu$.
It is possible that the viscosities used (especially $\nu = \scinot{2}{-6}$) are close to the intrinsic numerical dissipation\footnote{We \emph{do not} call this `numerical viscosity', since there is no reason to believe that the dissipation acts exactly like a physical viscosity} of \textsc{Fargo}.
However, even the difference between the $\nu = 10^{-5}$ and $\nu = \scinot{9}{-6}$ cases seems to be less than that predicted.
Unfortunately, attempting to run with viscosities higher than $10^{-5}$ was not reliably possible with \textsc{Fargo}.

\section{Conclusion}
\label{sec:conclude}

In this paper, we have continued the work of \citet{2007ApJ...663.1325E}, examining the migration rate of gap-opening planets.
We have shown that the migration rate is inversely proportional to the planet's mass, as previously predicted.
This strongly suggests that Type II migration should be regarded as an angular momentum exchange between the planet and disc, rather than the planet merely acting as a `relay station' between the inner and outer discs.

However, we have not seen the migration rate following the expected variation with disc viscosity.
Our numerical experiments showed a much weaker variation than that predicted.
It is possible that this is a numerical artifact; the intrinsic numerical dissipation of \textsc{Fargo} may be comparable to that produced by the physical viscosity we introduced.
The best way of testing this would be to repeat our experimental set up using a different hydrodynamics code.
We have no particular reason to believe that \textsc{Fargo} is misbehaving, but the possibility cannot be ruled out without an independent test.

Type II migration is not as simple a process as often thought.
The migration rate depends strongly on the the masses of both the disc and the planet, and this should be accounted for in monte carlo models of planetary system formation.


\bibliography{general}
\bibliographystyle{astron}


\section*{Acknowledgements}

The author acknowledges support from NSF grants AST-0406799, AST-0098442, AST-0406823, and NASA grants ATP04-0000-0016 and NNG04GM12G (issued through the Origins of Solar Systems Program).

\end{document}